\documentclass[12pt]{article}
\usepackage{newtxtext,newtxmath}
\usepackage{graphicx}
\usepackage[letterpaper,margin=1in]{geometry}
\renewenvironment{abstract}
	{\quotation}
	{\endquotation}
\date{}

\makeatletter
\renewcommand{\fnum@figure}{\textbf{Figure \thefigure}}
\renewcommand{\fnum@table}{\textbf{Table \thetable}}
\makeatother
\usepackage{scicite}
\usepackage{url}
\def\scititle{Flexibility as a Universal Nature-Inspired Mechanism for Thrust Enhancement}
\title{\bfseries \boldmath \scititle}
\author{
Roberta~Santoriello$^{1}$,
Francesco~Viola$^{2,3}$,
Vincenzo~Citro$^{1,4}$\and
\small$^{1}$DIIN, University of Salerno, Via Giovanni Paolo II, 84084 Fisciano, Italy.\and
\small$^{2}$Gran Sasso Science Institute, Viale F. Crispi 7, 67100 L'Aquila, Italy.\and
\small$^{3}$INFN--Laboratori Nazionali del Gran Sasso, 67100 Assergi, Italy.\and
\small$^{4}$CNR--Nanotec, unit\`a di Roma, P.le Aldo Moro 5, 00185 Rome, Italy.\and
\small$^\ast$Corresponding author. Email: vcitro@unisa.it\and
}
\begin{document} 
\maketitle
\vspace{-1.0cm}
\begin{abstract} \bfseries \boldmath
Nature has equipped jet-propelled swimmers with flexible nozzles that outperform rigid ones, yet the origin of this advantage has remained unexplained.
By tracking \emph{where} and \emph{when} energy is exchanged between fluid and structure, three-dimensional numerical simulations resolve the underlying mechanism: a standing-wave response of the nozzle, in which the structure dilates and then recoils synchronously, charging and releasing energy to enhance thrust.
Outside of this regime, the structure exhibits a traveling wave response, with expansion and contraction coexisting along the nozzle, reducing the thrust gain.
We propose a physics-based model that captures the boundary between standing and traveling responses in a closed form, showing that the optimum occurs when the natural period of the structure matches the pulse duration.
\textcolor{black}{Beyond this optimum the strain imposed by the nozzle curvature required for steering selects the geometry observed across marine species.}
The propulsion and maneuverability are reconciled within a single framework that yields design principles for soft robotic propulsors.
\end{abstract}


\noindent
\section*{INTRODUCTION}

\begin{figure*}
\centering
\includegraphics[width=\linewidth]{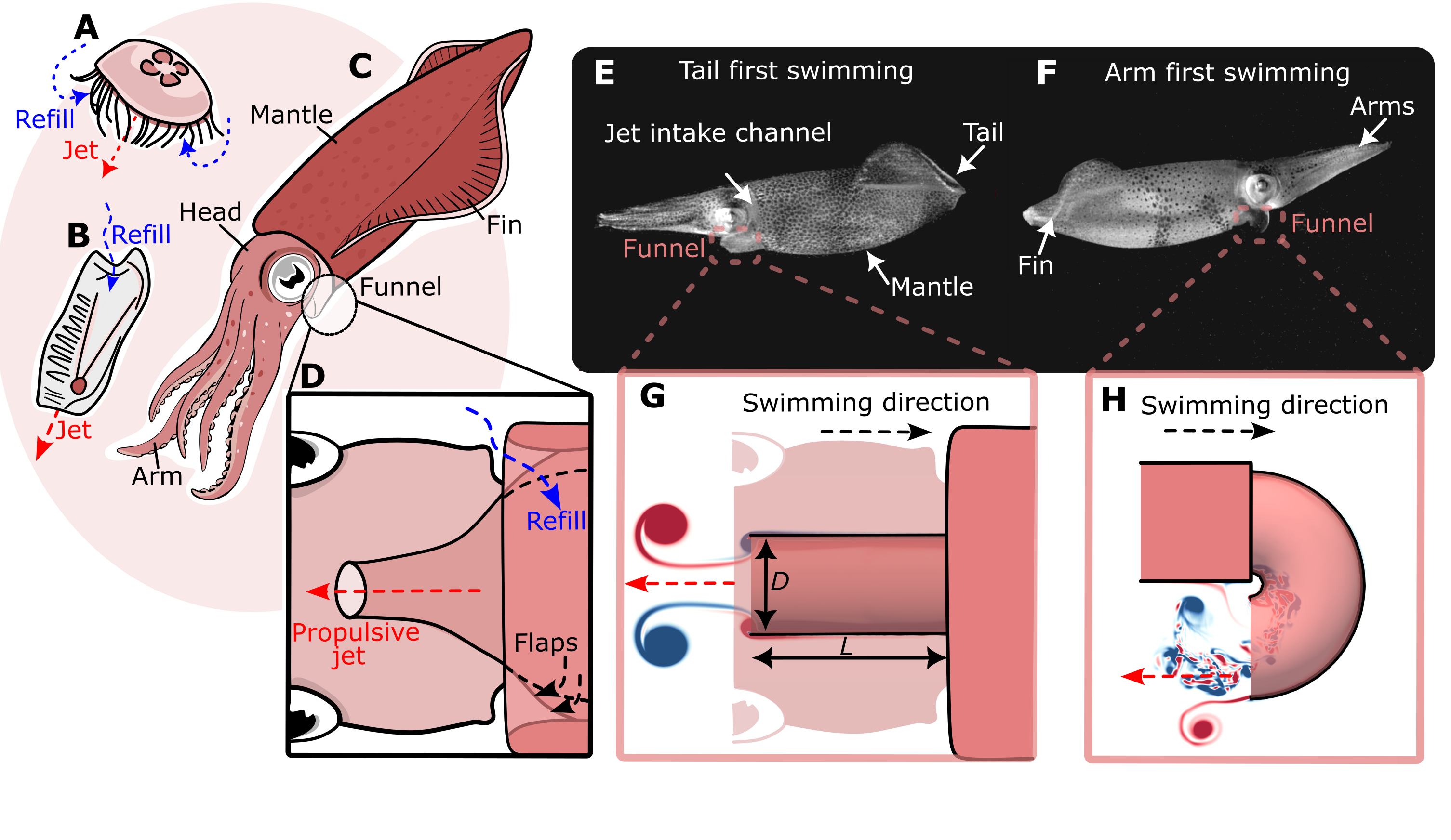}
\caption{\small \textbf{From straight to curved nozzles.}
(\textbf{A}-\textbf{B})~Sketch of (\textbf{A}) a jellyfish and (\textbf{B}) a salp, with blue arrows marking the refilling inflow and red arrows the direction of the propulsive jet.
(\textbf{C}-\textbf{D})~Schematic of (\textbf{C}) a squid and (\textbf{D}) its jet-propulsion cycle. During the refill phase (blue arrow) the mantle expands and water enters the cavity through lateral openings near the head.
The mantle then contracts, sealing the inlets (black arrow) and expelling the water as a propulsive jet (red arrow) through the funnel.
(\textbf{E}-\textbf{F})~The brief squid \textit{Lolliguncula brevis} swimming (\textbf{E}) tail-first and (\textbf{F}) arms-first [adapted with permission from \cite{bartol2016volumetric}].
(\textbf{G}-\textbf{H})~Our model, a thin compliant nozzle of diameter $D$ and length $L$, in (\textbf{G}) tail-first and (\textbf{H}) arms-first configurations.
\textcolor{black}{The dashed black arrows indicate the swimming direction from left to right.}}
\label{fig1}
\end{figure*}

Pulse-jet swimmers, such as jellyfish, salp and squid, propel themselves through cyclic contraction and relaxation of a flexible cavity \cite{gemmell2021cool}.
In each of them, the same fill-and-expel scheme recurs, with fluid drawn into the cavity during relaxation and ejected during contraction.
A jellyfish uses its bell for both phases, drawing water in and expelling it through the same opening
\cite{demont1988mechanics,dabiri2005flow,costello2021hydrodynamics} (\textcolor{black}{Fig.\ref{fig1}\textbf{A}}), a salp draws water in through an anterior oral siphon and ejects it through a posterior atrial one
\cite{bone1983jet,sutherland2010comparative} (\textcolor{black}{Fig.\ref{fig1}\textbf{B}}), while a squid draws water in through lateral openings flanking the head and expels it through a funnel \cite{gosline1985jet,kier2003muscle,bartol2001swimming} (\textcolor{black}{Figs.\ref{fig1}\textbf{C-D}}).
That such similar solutions have arisen independently across cnidarians, chordates and cephalopods points to a deeper physical principle shared by jet-propelled swimmers.

Among these swimmers, squids have evolved a particularly refined version of the jet-propulsion swimming mode \cite{anderson2005jet}, that has deep evolutionary roots.
Squids are believed to descend from monoplacophoran-like molluscs, with the gradual internalization and reduction of the conical external shell being a key step in their diversification towards active modes of life and increasingly complex behavior~\cite{kroger2011cephalopod}.
As the external shell was reduced, the ancestral foot was progressively remodeled into the funnel~\cite{shigeno2010origins}, a collagen-rich compliant nozzle \cite{gosline1983role} that the animal can orient enabling locomotion in two distinct modes~\cite{bartol2001swimming,bartol2016volumetric}: tail-first (\emph{backward}) swimming (\textcolor{black}{Fig.\ref{fig1}\textbf{E}}), with the posterior closed end of the mantle and the fins leading, and arms-first (\emph{forward}) swimming (\textcolor{black}{Fig.\ref{fig1}\textbf{F}}), with the arms leading and the fins and mantle trailing.
During routine cruising and maneuvering (often termed Mode~I swimming), squids propel through short pulses that expel only a small fraction of the mantle volume \cite{bartol2001swimming,anderson2000mechanics,krueger2009vortex}.
Each pulse rolls up into an isolated vortex ring that entrains a large volume of ambient fluid and thereby attains high propulsive efficiency \cite{bartol2016volumetric}.
This pulse-jet mechanism has been canonically reproduced in the laboratory through a piston-driven facility \cite{anderson2005jet} that expels a finite slug of fluid through a cylindrical rigid nozzle and the ejected fluid rolls up into a vortex ring \cite{saffman1975formation,gharib1998universal,dabiri2009optimal}.
Because the stroke is finite, the flow inside the nozzle decelerates once ejection ends, distinguishing the pulsed jet from a continuously fed one \cite{choi2022flow,mitchell2025formation}.

However, in most of the literature, the jet is expelled through a rigid nozzle \cite{gharib1998universal,anderson2005jet}.
The squid funnel, by contrast, is flexible, recalling a long-standing theme in cephalopod biology \cite{bujard2021resonant}.
Nearly half a century ago, Gosline and Shadwick \cite{gosline1983role} showed that the collagen fibers of the squid mantle act as passive springs, storing the work performed by the circular muscles and subsequently releasing it to refill the cavity \cite{macgillivray1999structure,krieg2012new}.

To isolate the role of flexibility in a controlled setting, Choi and Park \cite{choi2024mechanism} replaced the canonical rigid nozzle with an elastic one.
In their experiments, a motor-driven piston ejects a fixed volume of water through a thin silicone nozzle into a quiescent water tank, while two-dimensional particle image velocimetry resolves the surrounding flow and a sub-pixel tracking technique records the deformation of the nozzle.
By varying the structural stiffness and the jet-exit velocity, they showed that flexibility produces a  thrust gain, increasing both the circulation and the hydrodynamic impulse of the jet relative to the rigid case.
Because this enhancement differed among the nozzles they tested, some flexibilities yielding larger thrust gains than others, they inferred the existence of an optimal flexibility, and conjectured that it may be governed by the same empirical principle that sets the optimum in a continuous jet \cite{choi2022flow}.
However, because flexibility was sampled at only three nozzles, one can identify only the best of the three rather than the \textcolor{black}{global optimum}.
Moreover, it is not guaranteed that the criterion established for the continuous jet \cite{choi2022flow} carries over to the pulsed one \cite{choi2024mechanism}; and, being empirical, this criterion leaves the physical mechanism behind the optimum opaque.
\textcolor{black}{The mechanism underlying the thrust gain is most likely energetic, as the experimental observations point to a flexible structure that stores and releases elastic energy \cite{choi2022flow,choi2024mechanism,mitchell2025formation}.
What remains to be resolved is \emph{where} on the nozzle and \emph{when} during the pulse this energy exchange occurs, since earlier analyses characterized the fluid--structure interaction in terms of space-averaged quantities \cite{choi2022flow}.}
\textcolor{black}{Resolving this local energetic mechanism would account for the optimal stiffness of a straight nozzle, yet the squid funnel also bends through nearly a hemisphere to steer the jet during arms-first swimming (Fig.\ref{fig1}\textbf{F}), where additional functional constraints may come into play beyond the hydrodynamic optimum.
Whether and how the thrust gain mechanism operates under the curvature required for steering (Fig.\ref{fig1}\textbf{H}) is so far unexplored, as existing studies \cite{choi2022flow,choi2024mechanism,mitchell2025formation} have focused only on straight nozzles.}

To address these points, we model the squid funnel as a flexible nozzle operating in both tail-first (\textcolor{black}{Figs.~\ref{fig1}\textbf{E-G}}) and arms-first (\textcolor{black}{Figs.~\ref{fig1}\textbf{F-H}}) configurations.
Our high-fidelity fluid--structure interaction (FSI) solver \cite{viola2023high} (see Materials and Methods for details) reproduces the behavior of highly flexible nozzles---which undergo buckling and collapse---as observed experimentally \cite{choi2024mechanism,mitchell2025formation}, and allows us to map the local power exchange between fluid and structure in space and time, making the store-and-release mechanism directly visible.
A simple model then identifies the physical origin of the thrust gain and predicts, in closed form, the nozzle stiffness at which it is maximized.
Extending the analysis to a curved nozzle, we show that the thrust gain persists under the curvature required for steering and that the strain imposed by bending selects the geometry observed across squid species.
Propulsion and maneuverability are thereby reconciled within a single physical framework that yields useful design principles for pulsed-jet soft swimmers.

\section*{RESULTS}

We model the squid funnel as a thin compliant nozzle of aspect ratio $\mathrm{AR}\text{=}L^*/D^*$ (\textcolor{black}{Fig.~\ref{fig1}\textbf{G}}), where $D^*$ and $L^*$ are the nozzle diameter and length, respectively.
As shown in \textcolor{black}{Fig.~\ref{fig2}\textbf{C}}, the jet is driven by a prescribed inflow velocity ($U_{\text{p}}$, also referred to as the piston velocity) consisting of an acceleration ramp \textcolor{black}{of duration} $T_{\text{acc}}$ (the time required for the inflow velocity to reach its maximum $U^*$) followed by a deceleration ramp \textcolor{black}{of duration} $T_{\text{dec}}$ (see Supplementary Materials for details).
Throughout this work, an asterisk marks dimensional quantities, which are made dimensionless using $D^*$, the fluid density $\rho_\text{f}^*$, and the maximum inflow velocity $U^*$.
To assess the effect of the nozzle flexibility on thrust gain, we perform a parametric sweep over the dimensionless flexibility  $K \text{=} E^* h^*/(\rho_\text{f}^* U^{*2} D^*)$, where $E^*$ and $h^*$ are the Young's modulus and thickness of the nozzle.
The sweep ranges from $K \text{=} 2$ to $K \text{=} \infty$, the latter denoting the rigid case.
The geometry is held fixed at $\text{AR} \text{=}2$ and the flow regime is characterized by the Reynolds number $Re \text{=} \rho_\text{f}^* U^* D^*/\mu^*\text{=}\text{4500}$ (with $\mu^*$ the dynamic viscosity), which sets the ratio of inertial to viscous forces.

Our simulation setup is inspired by the experiments of Choi and Park \cite{choi2024mechanism}, and only to facilitate a direct comparison with their data, we introduce the two dimensionless parameters they adopt to characterize the system (elsewhere in the paper we retain the non-dimensionalization based on $U^*$).
Let $u_{\text{m,r}}^*$ denote the maximum centerline velocity at the nozzle exit in the rigid case, reached at time $T_{\text{acc,out}}^*$, \textcolor{black}{(which slightly lags the inflow acceleration time $T_{\text{acc}}$ owing to the jet development along the nozzle; see Fig.\ref{fig:S1}\textbf{E} in the Supplementary Materials)}.
The effective acceleration time  $\tau \text{=} u_{\text{m,r}}^* T_{\text{acc,out}}^*/L^*$ weighs the jet acceleration time $T_{\text{acc,out}}^*$ against the transit time $L^*/u_{\text{m,r}}^*$ that the jet takes to cross the nozzle, and thus measures how impulsively the stroke is delivered.
The effective stiffness $K_{\text{eff}}\text{=}E^* h^*/(\rho_\text{f}^* u_{\text{m,r}}^{*2} D^*)$, built on the rigid-case exit velocity, represents the ratio of the tensile stress in the flexible nozzle to the fluid pressure acting on the nozzle cross-sectional plane \cite{choi2022flow,choi2024mechanism}.

\textcolor{black}{To characterize how the jet responds as $K$ varies, we follow previous studies \cite{choi2022flow,choi2024mechanism,mitchell2025formation} and monitor three quantities, each normalized by the corresponding value in the rigid-nozzle case.}
The jet centerline exit-velocity gain is defined as $\hat{u}_\text{e}\text{=}u_{\text{m,K}}^*/u_{\text{m,r}}^*$, where $u_{\text{m,K}}^*$ is its value at stiffness $K$ and $u_{\text{m,r}}^*$ the rigid-nozzle counterpart.
The circulation gain is defined as $\hat{\Gamma}\text{=}\Gamma/\Gamma_\text{r}$, where $\Gamma$ is the circulation at stiffness $K$ and $\Gamma_\text{r}$ the rigid-nozzle counterpart, both computed as $\Gamma \text{=}\int_{S^+}\omega_x\,\mathrm{d}A$; here $\omega_x$ is the out-of-plane component of the vorticity vector $\boldsymbol{\omega}$, and $S^+$ is the half-plane through which the vortex ring propagates, extending one and a half diameters in the transverse direction ($y$) and eight diameters downstream of the nozzle exit.
To measure the thrust generated by the vortex ring \cite{saffman1992vortex,bartol2009pulsed,choi2024mechanism}, we evaluate the hydrodynamic impulse $\boldsymbol{I}_\text{h}\text{=}0.5\,\rho_\text{f}\int_V (\boldsymbol{x}\times\boldsymbol{\omega})\,\mathrm{d}V$ over a rectangular control volume $V\text{=}\{\,|x|,|y|\le 2D,\ 2D\le z\le 7.5D\,\}$ downstream of the nozzle exit.
We report its magnitude $I_\text{h}\text{=}\lVert\boldsymbol{I}_\text{h}\rVert$, normalized by the corresponding rigid-nozzle value $I_{\text{h,r}}\text{=}\lVert\boldsymbol{I}_{\text{h,r}}\rVert$, as $\hat{I}_\text{h}\text{=}I_\text{h}/I_{\text{h,r}}$, and we refer to this ratio as the thrust gain.
In what follows, both circulation and impulse are evaluated at $t\text{=}4$.

Beyond this straight configuration, we also consider a curved one (Fig.\ref{fig1}\textbf{H}), in which the nozzle is bent by 180 degrees to mimic arms-first swimming (Fig.\ref{fig1}\textbf{F}).
The straight funnel (Fig.\ref{fig1}\textbf{G}) exposes the local mechanism by which flexibility amplifies thrust, while the curved one (Fig.\ref{fig1}\textbf{H}) tests how that mechanism changes under the curvature required for steering.
We turn first to the straight case.

\subsection*{Energy exchange in the straight nozzle}

\begin{figure*}
\centering
\includegraphics[width=\linewidth]{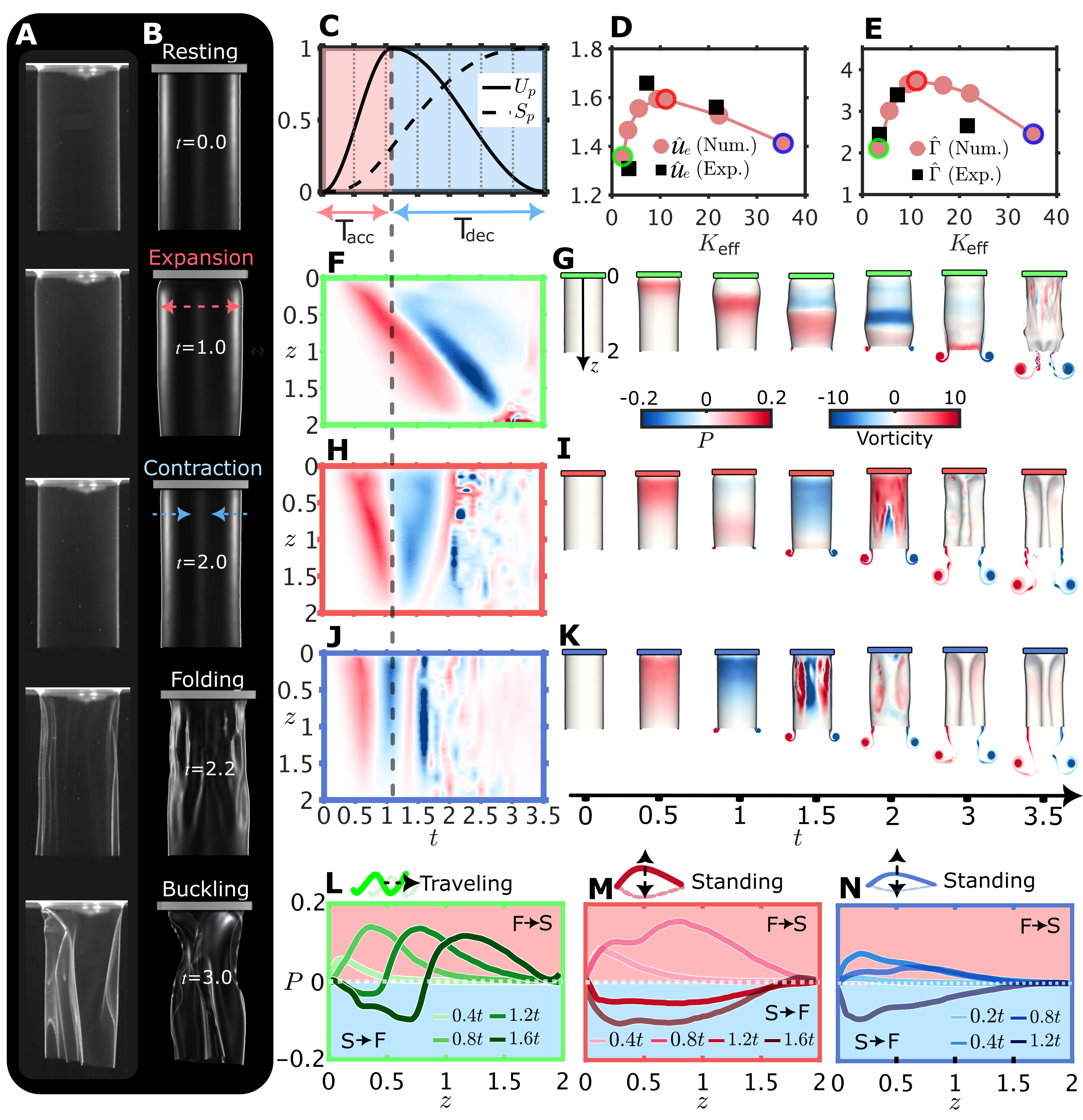}
\caption{\footnotesize \textbf{Energy exchange.} (\textbf{A--B}) Flexible nozzle at five characteristic instants during the pulse: (\textbf{A}) experiment [adapted from \cite{choi2024mechanism}] and (\textbf{B}) numerical simulations ($\tau \text{=} 0.48$, $K \text{=} 8$).
(\textbf{C}) Piston velocity ($U_\text{p}$) and displacement ($S_\text{p}$, normalized by its maximum) versus time; red and blue regions mark the acceleration ($T_\text{acc}$) and the deceleration ($T_\text{dec}$) phases.
(\textbf{D}) Jet centerline exit-velocity and (\textbf{E}) circulation gains versus $K_{\text{eff}}$: present simulations (circles), experiment of \cite{choi2024mechanism} (squares); green- and blue-outlined markers denote the two suboptimal cases, red the optimal one.
(\textbf{F}, \textbf{H}, \textbf{J}) Spatiotemporal evolution of the local power along a nozzle generatrix and (\textbf{G},\textbf{I},\textbf{K}) nozzle colored by the local power ($P$) with the surrounding fluid colored by the out-of-plane vorticity on the center-plane; (\textbf{F},\textbf{G}) $K \text{=} 2$, (\textbf{H},\textbf{I}) $K \text{=} 10$ and (\textbf{J},\textbf{K}) $K \text{=} 32$.
(\textbf{L--N}) Local power along a nozzle generatrix at four instants, with F$\rightarrow$S (red background) marking the fluid charging the solid and S$\rightarrow$F (blue background) the solid releasing energy: (\textbf{L}) $K \text{=} 2$ (traveling-wave), (\textbf{M}) $K \text{=} 10$ (standing-wave, optimal case) and (\textbf{N}) $K \text{=} 32$ (standing-wave).}
\label{fig2}
\end{figure*}

As shown in Fig.\ref{fig2}\textbf{B}, our three-dimensional FSI simulations recover the structural response observed experimentally \cite{choi2024mechanism} and shown in Fig.\ref{fig2}\textbf{A}, which unfolds in four successive phases \cite{choi2024mechanism,mitchell2025formation}.
Starting from rest, the fluid accelerates following the velocity history of Fig.\ref{fig2}\textbf{C}, generating a positive internal pressure that drives the flexible structure outward (second snapshot of Fig.\ref{fig2}\textbf{A} and Fig.\ref{fig2}\textbf{B}).
The nozzle then returns towards its undeformed configuration (third snapshot) and is subsequently drawn inward past it (fourth snapshot).
At later times, the structure buckles into a folded configuration (fifth snapshot).
The simulations resolve the full three-dimensional kinematics of this inward-collapse pattern and the buckled geometry observed experimentally \cite{choi2024mechanism,mitchell2025formation}, extending prior numerical studies of flexible nozzles \cite{singh2026elastic} into the high-compliance regime in which buckling emerges.

Figure \ref{fig2}\textbf{D} reports the jet centerline exit-velocity gain ($\hat{u}_\text{e}$) as a function of the effective stiffness ($K_{\text{eff}}$), for both the present simulations (circular markers) and the experiments of \cite{choi2024mechanism} (square markers).
The simulations follow the experimental trend, with $\hat{u}_\text{e}$ increasing with $K_{\text{eff}}$ up to a peak and decreasing thereafter, so that the maximum value of $\hat{u}_\text{e}$ is attained at an intermediate stiffness.
In Fig.~\ref{fig2}\textbf{E}, the same non-monotonic behavior is observed for the circulation gain ($\hat{\Gamma}$), which peaks at the same intermediate stiffness as $\hat{u}_\text{e}$.
In what follows, we refer to these peaks as ``optimal'' conditions and seek to establish the physical origin of this enhancement.

It has been proposed \cite{choi2024mechanism} that the underlying mechanism may be analogous to that of a flexible nozzle ejecting a continuous jet \cite{choi2022flow}.
For that configuration, the optimal flexibility was identified empirically as the stiffness maximizing the jet centerline exit-velocity, circulation, and impulse gains, with the optimum reached at a dimensionless wave speed $\hat{c}\text{=}\tau\sqrt{K_{\text{eff}}}\simeq 3$.
However, because flexibility was sampled at only three nozzles, these experiments identify the best of the three rather than the global optimum.
Moreover, it is not guaranteed that the criterion established for the continuous jet \cite{choi2022flow} carries over to the pulsed one \cite{choi2024mechanism}.
Indeed, both the experimental data and the numerical results in Figs.~\ref{fig2}\textbf{D-E} show that $\hat{u}_\text{e}$ and $\hat{\Gamma}$ peak at $K_{\text{eff}}\approx 10.67$, corresponding to $\hat{c}\approx 1.56$, well below the optimal value $\hat{c}\simeq 3$ predicted by the continuous-jet criterion \cite{choi2022flow}.

These observations suggest that a deeper physical mechanism remains to be uncovered.
To unveil its origin, we turn to a local energetic description of the FSI and compute, for the first time, the power exchanged between the fluid and the inner surface of the flexible nozzle,
\begin{equation}
P(\boldsymbol{x},t) = p\textcolor{black}{(\boldsymbol{x},t)}\,v_n\textcolor{black}{(\boldsymbol{x},t)},
\label{eq:local_power}
\end{equation}
where $\boldsymbol{x}\text{=}(x,y,z)$ is a point on the inner surface of the nozzle, $p(\boldsymbol{x},t)$ is the local pressure on the inner surface of the structure, and $v_n(\boldsymbol{x},t)$ is the \textcolor{black}{wall-normal velocity of the nozzle}, positive when the structure expands.
With this convention, $P>0$ identifies regions where the fluid charges the structure, and $P<0$ regions where the structure returns energy to the fluid.

The local power field ($P$) defined in Eq. \eqref{eq:local_power} reveals \emph{where} and \emph{when} the thrust gain mechanism operates and shows that, depending on $K$, the structural response organizes into distinct regimes.
Figure \ref{fig2}\textbf{F} reports the spatiotemporal evolution of $P$ along a generatrix of the cylinder for a highly flexible case ($K\text{=}2$, green marker in Figs.\ref{fig2}\textbf{D-E}), while Fig.\ref{fig2}\textbf{G} displays $P$ on the nozzle surface itself, together with the structural deformation and the surrounding flow field colored by the out-of-plane vorticity.
Starting from rest, the fluid charges the compliant structure, driving its radial expansion ($t\text{=}0.5$ and $t\text{=}1$ in Fig.\ref{fig2}\textbf{G}).
The nozzle then begins to discharge, returning energy to the fluid near the inlet, while the fluid continues to charge the structure near the exit ($t\text{=}1.5$ and $t\text{=}2$), so that charging and discharging coexist along the nozzle.
In the spatiotemporal map of Fig.\ref{fig2}\textbf{F}, this coexistence appears as oblique bands in the $(z,t)-$plane.
The discharge then propagates downstream and eventually spreads over the entire structure ($t\text{=}3$), until the structure collapses through buckling ($t\text{=}3.5$).
This spatiotemporal pattern is the signature of a traveling-wave regime: the power propagates downstream along the generatrix, as further illustrated in Fig.\ref{fig2}\textbf{L}, which shows the local power profiles at successive instants and along the cylinder generatrix.

Fig.\ref{fig2}\textbf{H} shows the same power-map for $K\text{=}10$, the case at which both $\hat{u}_\text{e}$ and $\hat{\Gamma}$ reach their peak (red marker in Figs.\ref{fig2}\textbf{D-E}).
The structural response now differs from the highly flexible case.
During the acceleration phase, the fluid charges the structure along its entire length, driving a uniform radial expansion ($t\text{=}0.5$ in Fig.~\ref{fig2}\textbf{I}).
The structure then discharges uniformly along its whole length as it relaxes back to its rest shape ($t\text{=}1.5$).
In the subsequent half-cycle the roles reverse.
The structure is charged inward ($t\text{=}2$) and reaches a buckled state ($t\text{=}2.5$).
Figure \ref{fig2}\textbf{M} confirms that the power distribution now resembles a standing wave rather than a traveling one: charging and discharging occur synchronously along the entire nozzle length, with no downstream propagation.
This standing-wave response persists for stiffer nozzles, such as $K\text{=}32$ in Figs.\ref{fig2}\textbf{J-K} (corresponding to the blue marker in Figs.\ref{fig2}\textbf{D-E}).
The dynamics mirrors the case $K\text{=}10$ but unfolds faster ($t\text{=}1.5$ and $t\text{=}2$ in Figs.\ref{fig2}\textbf{J-K}), as expected from the larger stiffness.
The generatrix map (Fig.\ref{fig2}\textbf{N}) confirms that all points move in unison.

\textcolor{black}{Taken together, these observations identify the optimal stiffness as the boundary between the traveling-wave and standing-wave regimes.}
The peaks in $\hat{u}_\text{e}$ (\textcolor{black}{Fig.~\ref{fig2}\textbf{D}}) and $\hat{\Gamma}$ (\textcolor{black}{Fig.~\ref{fig2}\textbf{E}}) \textcolor{black}{occur at this boundary, where the structure responds as a standing wave, storing energy from the fluid during the acceleration phase and releasing it back during deceleration.}
This is evident from Figs.~\ref{fig2}\textbf{C} and \textbf{H}.
For $t<T_{\text{acc}}$ the power map shows only a red region, indicating that the fluid charges the structure, whereas for $t>T_{\text{acc}}$ the map turns blue and the structure returns energy to the fluid until buckling sets in.
In the traveling-wave regime (Fig.~\ref{fig2}\textbf{F}), by contrast, charging and releasing coexist at different stations along the nozzle length at the same instant.
In this case a thrust gain is still produced, but it is smaller than in the standing-wave case.

\subsection*{Universality across nozzle lengths}

\begin{figure*}
\centering
\includegraphics[width=\linewidth]{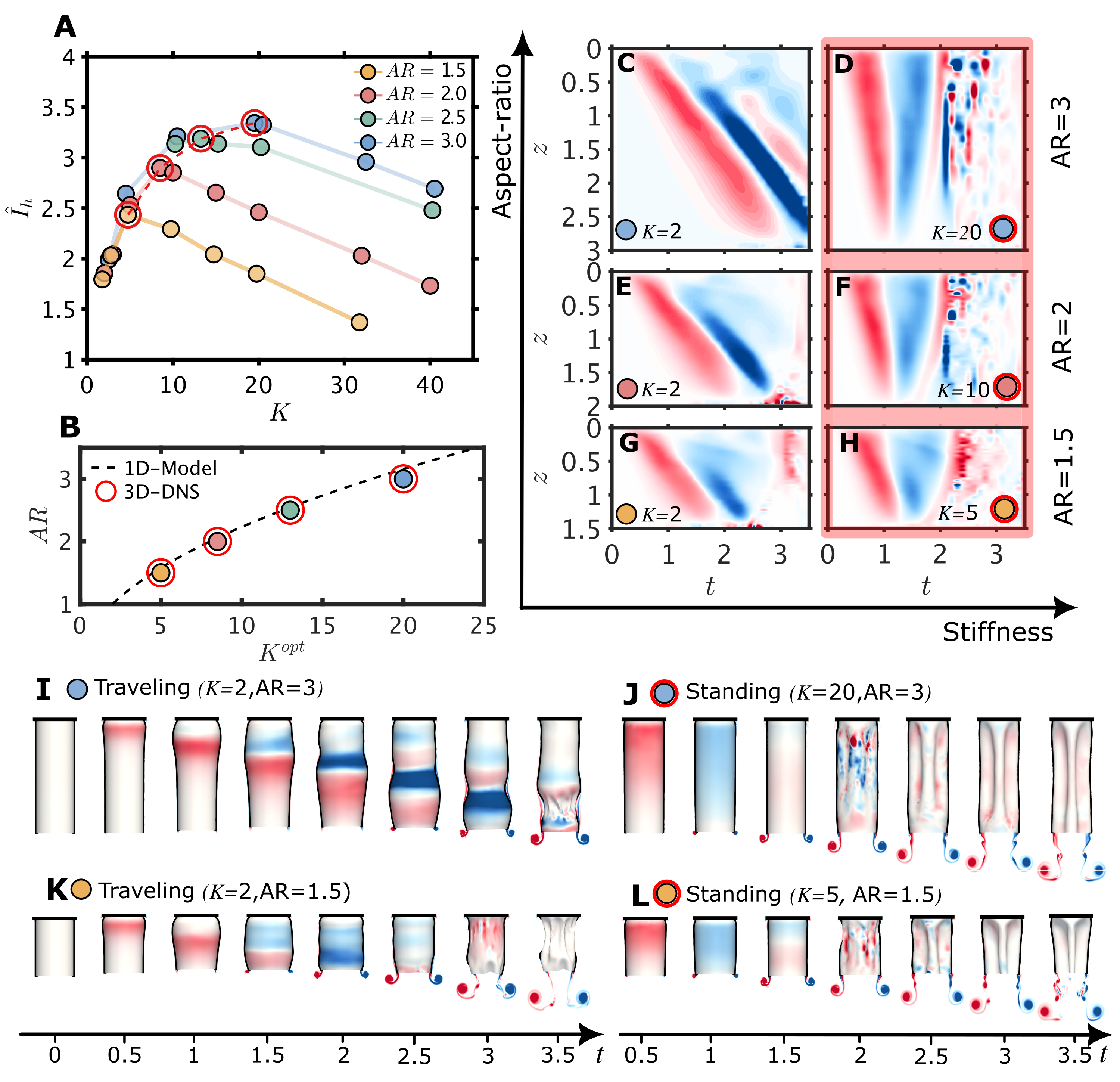}
\caption{\footnotesize \textbf{Energy exchange across nozzle aspect ratios and model predictions.}
(\textbf{A}) Hydrodynamic impulse for different aspect ratio ($AR$) as a function of $K$.
(\textbf{B}) $AR$ versus the optimal stiffness ($K^{\mathrm{opt}}$) predicted by the numerical simulations (red markers) and by the physics-based model (dashed black line).
(\textbf{C}--\textbf{H}) Local power for nozzles of different $AR$ and $K$: (\textbf{C}) $K\text{=}2$, $AR\text{=}3$, (\textbf{D}) $K\text{=}20$, $AR\text{=}3$, (\textbf{E}) $K\text{=}2$, $AR\text{=}2$, (\textbf{F}) $K\text{=}10$, $AR\text{=}2$, (\textbf{G}) $K\text{=}2$, $AR\text{=}1.5$ and (\textbf{H}) $K\text{=}5$, $AR\text{=}1.5$.
(\textbf{I}--\textbf{L}) Power on the flexible nozzle together with the surrounding flow field, colored by the out-of-plane vorticity: (\textbf{I}) $K\text{=}2$, $AR\text{=}3$, (\textbf{J}) $K\text{=}20$, $AR\text{=}3$, (\textbf{K}) $K\text{=}2$, $AR\text{=}1.5$ and (\textbf{L}) $K\text{=}5$, $AR\text{=}1.5$.
The color-map for the power and the vorticity field is the same as in Fig.~\ref{fig2}.}
\label{fig3}
\end{figure*}

\textcolor{black}{The optimum thrust-enhancement criterion for a flexible nozzle unraveled above sets the optimal stiffness as the threshold at which the structural response transitions from a traveling-wave to a standing-wave regime, in phase with the inflow velocity.}
To assess the robustness of this thrust gain mechanism, we repeat the energetic analysis for nozzles of different aspect ratios, $\mathrm{AR}\text{=}1.5,2,2.5,3$, while keeping the inflow conditions constant.
For every $\mathrm{AR}$ and $K$ considered, we evaluate the thrust gain generated by the vortex ring through the hydrodynamic impulse normalized by its corresponding rigid-nozzle case value (\textcolor{black}{Fig.\ref{fig3}\textbf{A}}).
In every case, $\hat{I}_h > 1$, so flexibility always enhances thrust relative to a rigid nozzle.
Each curve in \textcolor{black}{Fig.\ref{fig3}\textbf{A}} exhibits a peak (red marker) at a stiffness $K^{\mathrm{opt}}$, which shifts towards larger values as the nozzle lengthens, that is, as $\mathrm{AR}$ increases.
The thrust gain rises steeply as $K$ approaches $K^{\mathrm{opt}}$ from below and reduces only gradually beyond it, so that the enhancement persists over a \textcolor{black}{broader} range on the stiff side ($K>K^{\mathrm{opt}}$) than on the soft side ($K<K^{\mathrm{opt}}$).
Below $K^{\mathrm{opt}}$ the structure operates in the traveling-wave regime.
This is evident in the power-maps of \textcolor{black}{Figs.\ref{fig3}\textbf{C},\textbf{E},\textbf{G}} for $\mathrm{AR}\text{=}3, 2, 1.5$, respectively, as well as in the snapshot sequences of \textcolor{black}{Figs.\ref{fig3}\textbf{I},\textbf{K}} ($\mathrm{AR}\text{=}3$ and $1.5$), which display the local power exchange on the nozzle surface together with the surrounding flow colored by the out-of-plane vorticity.
In this regime, charging and discharging coexist \textcolor{black}{along the span of the nozzle}, with the fluid charging the structure in some regions while the structure returns energy to the fluid in others.

At $K^{\mathrm{opt}}$ the structure enters the standing-wave regime.
\textcolor{black}{Figs.\ref{fig3}\textbf{D},\textbf{F},\textbf{H}} ($\mathrm{AR}\text{=}3,2, 1.5$ at their peak stiffnesses $K^{\mathrm{opt}}\text{=}20, 10, 5$) show that charging and discharging now occur synchronously along the entire nozzle length, with every point expanding and recoiling in phase.
The same behavior is visible in the power-maps of \textcolor{black}{Figs.\ref{fig3}\textbf{J-L}}, where the nozzle is shown together with the surrounding flow.
Across all aspect ratios, the fluid charges the structure at $t\text{=}0.5$, the structure returns energy to the jet at $t\text{=}1$ and the inward re-charge at $t\text{=}1.5$ marks the onset of buckling (see $t>1.5$).

Two aspects of this picture are central.
First, in the standing-wave regime the local power exchange is spatially coherent.
At any instant of the pulse, the whole structure charges or releases energy with the same sign, which is the hallmark of the thrust gain mechanism.
Second, charging and releasing are tightly synchronized with the acceleration and deceleration phases of the inflow velocity, so that fluid charges the structure during the acceleration phase, while the structure returnes energy to the fluid during the deceleration phase, reinforcing the vortex ring.

These two features account for the dependence of $\hat{I}_\text{h}$ on the nozzle length.
In the standing-wave regime ($K \ge K^{\mathrm{opt}}$), the entire structure acts in phase, so a longer nozzle presents a larger area contributing synchronously to the jet, and the $\hat{I}_\text{h}$ curves are well separated in \textcolor{black}{Fig.\ref{fig3}\textbf{A}}.
In the traveling-wave regime, by contrast, only the portion of the structure spanned by the traveling wave is active at any instant, irrespective of the nozzle length, and the curves for different aspect ratios collapse onto one another.
Therefore, the length scale that sets the thrust is the wavelength in the traveling regime, largely insensitive to the nozzle length, and the full nozzle length in the standing regime.

The dependence of $K^{\mathrm{opt}}$ on nozzle length (red markers in Fig.\ref{fig3}\textbf{A}) can be explained by considering the flexible nozzle as a clamped--free elastic tube whose radial deformation obeys the Moens--Korteweg (MK) wave equation \cite{choi2022flow}, with characteristic speed $c\text{=}\sqrt{K}$ [equivalently $c^*\text{=}\bigl(E^*h^*/(\rho_\text{f}^*\,D^*)\bigr)^{1/2}$, in dimensional form].
The eigenproblem associated with the MK equation admits a discrete sequence of standing-wave modes, whose fundamental is a quarter-wave, with a node at the clamped inlet, a maximum at the free outlet, and period $T_1\text{=}4L/c$ (see Supplementary Materials for details).
Its shape resembles the spatial signature of the in-phase dilation-recoil observed in the numerical simulations when the wall response lies in the standing-wave regime.
The transition from the traveling-wave to the standing-wave regime is selected by a timing condition.
The fundamental period of the structure $T_1$ must match an effective pulse duration $T_\text{p}$, obtained from the acceleration and deceleration frequencies of the piston (see Supplementary Materials for details), i.e.\ $T_1\text{=}T_\text{p}$, giving the optimal stiffness in closed form,
\begin{equation}
K^{\text{opt}}(L, T_\text{p}) = 16  \left(\frac{L}{T_\text{p}}\right)^{2}.
\label{eq:Kstar}
\end{equation}
Figure \ref{fig3}\textbf{B} compares the optimal stiffness $K^{\mathrm{opt}}$ predicted by Eq.~\eqref{eq:Kstar} with that obtained from the 3D simulations for $\mathrm{AR} \text{=} 1.5,\,2,\,2.5,\,3$, showing good agreement across the full range and demonstrating the $K^{\text{opt}}\propto L^2$ scaling suggested by the numerical simulations, without any fitting constant.
The condition~(\ref{eq:Kstar}) marks the boundary between the two distinct regimes: a standing-wave response, in which expansion and contraction take turns over the entire structure, and a traveling-wave response, in which they coexist along the nozzle length and partly cancel.
\textcolor{black}{Moreover it provides a useful design rule for soft pulsed-jet propulsors, fixing the optimal stiffness ($K^{\text{opt}}$) from the nozzle length ($L$) and pulse duration ($T_{\text{p}}$).}

\subsection*{\textcolor{black}{Energy exchange in curved nozzles: structural bending constraint}}

During arms-first swimming (Figs.\ref{fig1}\textbf{F}--\textbf{H}), the squid funnel is bent through nearly an hemisphere beneath the body \cite{bartol2001swimming,bartol2016volumetric}.
To assess whether the thrust gain mechanism identified in the previous section persists under such curvature, we repeat the numerical analysis on a curved nozzle with $\text{AR=}2$, holding the same inflow conditions as in the straight case.

\begin{figure*}
\centering
\includegraphics[width=\linewidth]{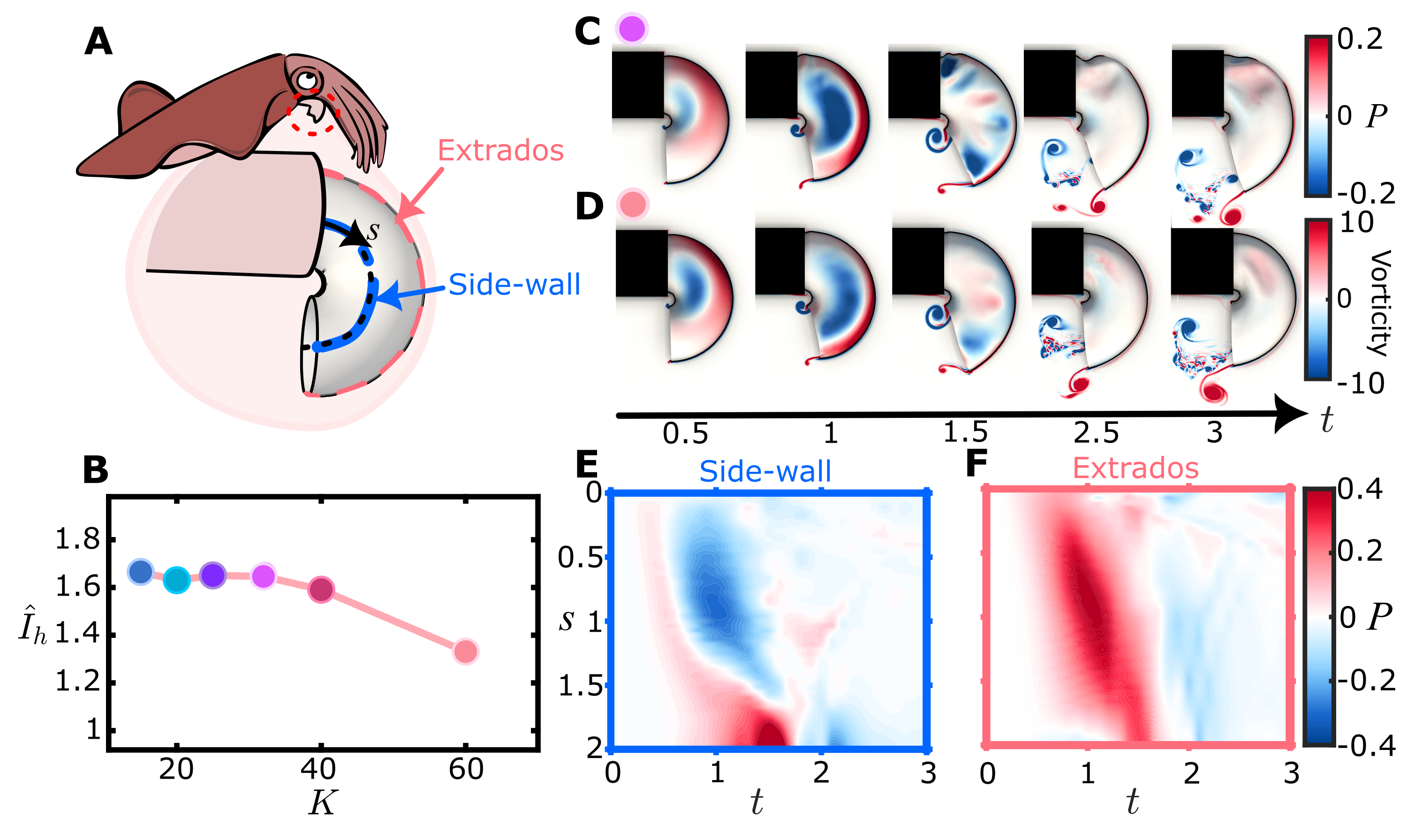}
\caption{\footnotesize \textbf{Energy exchange in the curved nozzle.}
(\textbf{A}) Schematics of a squid during arms-first swimming and  the curved nozzle, indicating the extrados and side wall.
(\textbf{B}) Normalized hydrodynamic impulse $\hat{I}_h$ versus $K$.
(\textbf{C}-\textbf{D}) Local power map on the curved nozzle at successive instants of the jet pulse, for (\textbf{C}) $K \text{=} 32$ and (\textbf{D}) $K \text{=} 60$; the nozzle is colored by the local power ($P$) and the flow field by the out-of-plane vorticity.
(\textbf{E}-\textbf{F}) Power maps for $K \text{=} 32$ along the (\textbf{E}) side wall and (\textbf{F}) extrados, as a function of arc length ($s$) along the nozzle axis and time ($t$).}
\label{fig4}
\end{figure*}

Figure~\ref{fig4}\textbf{B} shows the magnitude of the hydrodynamic impulse, normalized by its rigid-nozzle value $(\hat{I}_h)$ as a function of $K$.
Unlike the straight case, the thrust gain remains nearly constant across the stiffnesses tested, decreasing by only $\sim 15\%$ at the stiffest nozzle ($K\text{=}60$).
Flexibility therefore still enhances thrust on the curved nozzle, but the mechanism behind the gain differs from the straight case.
Figures~\ref{fig4}\textbf{C} and \textbf{D} show the local power exchange at successive instants for $K\text{=}32$ and $K\text{=}60$, respectively.
The nozzle response now varies around the \textcolor{black}{nozzle span}, differing between the outer side of the nozzle (extrados) and the side walls (see the schematic in Fig.\ref{fig4}\textbf{A}), in contrast to the axisymmetric dilation--recoil pattern of the straight case.
During the acceleration phase ($t\text{=}0.5$), the fluid charges the extrados outward, while the side walls return energy to the fluid as they recoil inward (Figs.~\ref{fig4}\textbf{C-D}).
At $t\text{=}1$, the release zones on the side walls grow and the charging region on the extrados shrinks, until the nozzle buckles at $t\text{=}1.5$ and the power distribution loses any spatial organization.

This mechanism can be seen more clearly in Figs.\ref{fig4}\textbf{E}--\textbf{F}, which show, for $K\text{=}32$, the local power exchange along a side wall generatrix (\textbf{E}) and along the extrados (\textbf{F}) generatrix.
Each generatrix is a longitudinal line traced on the wall at a fixed azimuthal position around the nozzle axis (Fig.~\ref{fig4}\textbf{A}).
In each power-map, the vertical axis is the arc length $s$ measured along the curved centerline of the nozzle, running from $s\text{=}0$ at the inlet to $s\text{=}L$ at the outlet, and the horizontal axis is the time $t$.
Both maps confirm that, at every instant, charging and releasing coexist along the structure, as a direct consequence of the curvature.
The standing-wave mode that, in the straight geometry, stored and released energy in phase with the pulse, no longer emerges.
Therefore, the thrust gain mechanism identified for the straight nozzle cannot operate on the curved geometry required for steering.
\textcolor{black}{The funnel geometry observed in nature is thus not set by this fluid--structure mechanism alone, but must also reflect functional constraints from maneuverability, since squid bend the funnel to steer the jet.}

This bending imposes a tensile strain on the structure that cannot exceed the maximum extensional strain ($\varepsilon_{\max}$) that the muscular tissue can sustain without failure.
Modeling the funnel as a thin compliant tube of radius $R^* \text{=} D^*/2$, curved along a circular arc of radius of curvature $R_\text{c}^*$, the extensional strain on the outer fiber is $\varepsilon \text{=} R^*/R_\text{c}^*$.
Combining with the arc-length relation $L^* \text{=} R_c^* \theta$ yields $\varepsilon \text{=} R^* \theta/L^*  \text{=} D^* \theta/(2 L^*)$, where $D^*$ is here identified with the funnel orifice diameter.
Imposing $\varepsilon \le \varepsilon_{\max}$ at the maximum deflection $\theta \text{=} \pi$ gives
\begin{equation}
AR=\frac{L^*}{D^*}\ge \frac{\pi}{2 \varepsilon_{\max}}.
\label{eq:LD-bound}
\end{equation}

To estimate the strain bound $\varepsilon_{\max}$ appearing in Eq.\eqref{eq:LD-bound}, we appeal to the closest characterized muscular hydrostats \cite{curtin2000energy,gosline1983role} which report a maximum strain of $\varepsilon_{\max}\sim 0.25$--$0.45$.
Substituting this range into Eq.~\eqref{eq:LD-bound}, gives, $AR \approx 3.5\text{--}6.28$ that closely aligns with the range $AR \in [2.8,5]$ observed across squid species (Fig.\ref{fig:S2}\textbf{C} in Supplementary Materials).
Although the strain limit is inferred by analogy and a uniform thin nozzle idealizes the funnel anatomy, this simple bound captures the observed aspect-ratio range, \textcolor{black}{showing that a functional requirement must be accounted for in setting the funnel geometry.
The same requirement may serve as a design guideline for soft swimmers, whose steerable propulsors must be slender enough to redirect the jet without exceeding the deformation their material can withstand.}

\section*{DISCUSSION}

By resolving the local fluid--structure energy exchange in a model squid funnel, we have identified the physical mechanism behind the optimal flexibility that enhances thrust: the structure executes an expansion--recoil cycle synchronized with the pulse, storing energy from the fluid during the acceleration phase and releasing it back during deceleration to reinforce the vortex ring.
This optimum holds across nozzle lengths and, in every case tested, marks the transition between a standing-wave regime, in which the whole nozzle breathes synchronously, and a traveling-wave regime, in which expansion and contraction coexist along the nozzle and partly cancel.
Our simulations agree with the experiments on flexible nozzles reported in the literature \cite{choi2024mechanism} and, beyond reproducing them, expose the physics behind the optimum through the local energy exchange we resolve.

A simple model then accounts for this optimum in closed form, recovering the $K^{\text{opt}}\propto L^2$ scaling suggested by the simulations and providing a useful design rule, free of any fitting constant, that fixes the optimal stiffness from the nozzle length and the pulse duration.
Together, these results show that flexibility can be harnessed for thrust enhancement, with a gain that persists across every case examined, optimal or not, relative to the rigid nozzle.

From a broader perspective, the standing-wave regime recalls the propulsion of other pulsed-jet swimmers, in particular salps, pointing to a shared underlying physical principle.
Indeed, salps are shaped like hollow tubes and swim through temporally coordinated contractions reminiscent of peristaltic pumping \cite{santoriello2026universality}, with the body contracting almost simultaneously \cite{sutherland2010comparative}.

Returning to the squid funnel, while the straight nozzle exposes the local mechanism by which flexibility amplifies thrust, the curved one tests how that mechanism operates under the curvature required for steering.
In the latter case, the nozzle response varies around the span, differing between the extrados and the side walls, in contrast to the axisymmetric expansion--recoil pattern of the straight case.
The curved configuration underscores that the funnel also supports maneuverability.
This function constrains the geometry, since the strain budget for bending the funnel to steer the jet sets a lower bound on its length, consistent with the aspect-ratio range observed across squid species.

Overall, this work shows how the principles of fluid--structure interaction can be translated into predictive models, deepening our understanding of the physical mechanism behind the flexibility exploited by marine organisms and yielding rational design principles for soft swimmers.
More broadly, it exemplifies a general lesson: the optimum is not the softest nozzle, but the one whose elastic clock runs at the pace of the stroke.

\newpage

\clearpage 
\bibliography{manuscript} 
\bibliographystyle{sciencemag}
\section*{Acknowledgments}

This work was supported by the National Recovery and Resilience Plan (NRRP), Mission 4, Component 2, Investment 1.1, Call for tender No. 1409 (published on 14.9.2022) by the Italian Ministry of University and Research (MUR), funded by the European Union - Project Title 2022J5ZNHS - Holes in the heart: a fluid dynamics approach to atrial septal defects – (CUP D53C24004100001, Grant Assignment Decree No. 20435 adopted on 08/11/2024) awarded to FV and VC;
and by the European Research Council (ERC) under the European Union’s Horizon Europe research and innovation program (Grant No. 101039657, CARDIOTRIALS) awarded to FV.

After finishing the work, we became aware of a related work recently posted on arXiv
\cite{choi2026squid}, which is complementary to the present study and addresses distinct aspects of
the problem.

\paragraph*{Author contributions:}

Conceptualization: RS, FV, VC.
Methodology: FV, VC.
Software: FV.
Validation: RS, FV, VC.
Formal Analysis: RS, FV, VC.
Resources: FV, VC.
Data curation: RS, FV, VC.
Writing—original draft: RS, FV, VC.
Writing—review \& editing: RS, FV, VC.
Visualization: RS.
Supervision: FV, VC.
Project Administration: FV, VC.
Funding acquisition: FV, VC.

\paragraph*{Competing interests:}

Authors declare that they have no competing interests.

\paragraph*{Data and materials availability:}

All data and code needed to evaluate and reproduce the results in the paper are present in the paper and/or the Supplementary Materials. No new materials have been created for this work.

\subsection*{Supplementary materials}
Materials and Methods\\
Supplementary Text\\
Figs. S1 to S2\\
Tables S1 \\

\newpage
\renewcommand{\thefigure}{S\arabic{figure}}
\renewcommand{\thetable}{S\arabic{table}}
\renewcommand{\theequation}{S\arabic{equation}}
\renewcommand{\thepage}{S\arabic{page}}
\setcounter{figure}{0}
\setcounter{table}{0}
\setcounter{equation}{0}
\setcounter{page}{1} 

\begin{center}
\section*{Supplementary Materials for\\ \scititle}
Roberta~Santoriello$^{\ast\dagger}$,
Francesco~Viola$^\dagger$,
Vincenzo~Citro\\ 
\small$^\ast$Corresponding author. Email: vcitro@unisa.it\\
\end{center}

\subsubsection*{This PDF file includes:}
Materials and Methods\\
Supplementary Text\\
Figures S1 to S2\\
Tables S1 \\

\newpage

\section*{Materials and Methods}

\subsection*{Fluid-structure interaction solver}

The numerical framework follows that of \cite{viola2023high} and consists of a flow solver two-way coupled with a structural one.
The fluid is governed by the incompressible Navier--Stokes equations, which in dimensionless form read,
\begin{equation}
\frac{\partial \boldsymbol{u}}{\partial t}+\boldsymbol{u}\cdot \nabla \boldsymbol{u} = -\nabla p+\frac{1}{Re}\nabla^2 \boldsymbol{u}+\boldsymbol{f},
\qquad
\nabla \cdot \boldsymbol{u} = 0,
\label{eq:NS}
\end{equation}
where $\boldsymbol{u}$ is the velocity field, $p$ the pressure and $\boldsymbol{f}$ an immersed-boundary volume forcing imposing the no-slip condition at the fluid--structure interface.
The equations are solved with a central second-order finite differences solver on a staggered Cartesian grid \cite{verzicco1996finite}.
For details of the numerical scheme the reader is referred to \cite{viola2023high}.
As described in \cite{viola2023high}, the flow solver provides the hydrodynamic loads ($\mathbf{F}_{\text{n}}^{\text{ext}}$) required to update the structural mechanics.

The deformation of the flexible nozzle is governed by a spring-network model based on an interaction-potential approach \cite{de2016moving}.
The wet surface is discretised into triangular elements with the body mass distributed uniformly across their vertices.
Adjacent nodes are connected by stretching springs of elastic constant $k_e$, and pairs of triangles sharing an edge are endowed with a bending stiffness $k_b$.
The elastic constants are computed using the model of \cite{gelder1998approximate}, $k_e \text{=} E h (A_1+A_2)/\ell_0^2$ and $k_b \text{=} 2B/\sqrt{3}$, with $E$ the Young's modulus, $B$ the bending stiffness, $h$ the membrane thickness, and $A_1,A_2$ the areas of the two triangles sharing the edge of rest length $\ell_0$.
The total elastic potential energy is $W \text{=} k_e(\ell-\ell_0)^2/2 + k_b[1-\cos(\vartheta-\vartheta_0)]$, with $(\ell_0,\vartheta_0)$ the reference stress-free edge length and dihedral angle. Internal forces $\boldsymbol{F}_{\rm int} \text{=} -\nabla W$, together with the external fluid forces $\boldsymbol{F}_{\rm ext}$, enter Newton's law at each node,
\begin{equation}
m_n\,\boldsymbol{\ddot{x}} = \boldsymbol{F}_{\rm int} + \boldsymbol{F}_{\rm ext}.
\label{eq:newton}
\end{equation}
Time integration of (\ref{eq:newton}) yields nodal velocities and positions, updating the structure configuration each step.
All simulations were performed with a GPU-parallelized solver \cite{viola2023high}.
Each run was distributed across two NVIDIA H100 GPUs and required approximately five hours of wall-clock time.

\newpage

\section*{Supplementary Text}

\subsection*{Imposed inflow}

In the following, dimensional quantities are denoted by an asterisk, and their non-dimensional counterparts are obtained by scaling with the nozzle diameter $D^*$, the maximum centerline inlet velocity $U^*$, and the fluid density $\rho_{\text{f}}^*$ (see Table~\ref{tab:S1}).
We recall the dimensionless groups introduced in the main text, namely the Reynolds number $\mathrm{Re}\text{=}\rho_{\text{f}}^* U^* D^*/\mu^*$ and the stiffness $K\text{=}E^* h^*/(\rho_{\text{f}}^* U^{*2} D^*)$.

In the numerical simulations, the jet is driven by a prescribed inflow velocity at the nozzle inlet that mimics the piston motion, $u(z\text{=} 0,r,t) \text{=}  U_{\text{p}}(t)\, f(r/R_0),$ which separates the temporal evolution $U_{\text{p}}(t)$ from the radial profile $f(r/R_0) \text{=} \tanh[\delta(1-r/R_0)]/\tanh\delta$ (with $\delta \text{=}  50$).
The profile is normalized to unity on the axis, $f(0)\text{=} 1$, so that $U_{\text{p}}(t)$ is the
instantaneous centerline velocity, normalized by its peak value $U^*$.

The kinematic of the piston $U_{\text{p}}(t)$ consists of a cosine acceleration ramp of duration $T_\mathrm{acc}$, a short plateau of duration $T_\mathrm{plateau}$, and a deceleration ramp of duration $T_\mathrm{dec}$,
\begin{equation}
U_{\text{p}}(t) =
\begin{cases}
\dfrac{U_{\text{m}}}{2} \left(1 - \cos\dfrac{\pi t}{T_\mathrm{acc}}\right), & 0 \le t < T_\mathrm{acc},\\[3mm]
U_{\text{m}}, & T_\mathrm{acc} \le t < T_\mathrm{acc}+T_\mathrm{plat},\\[3mm]
\dfrac{U_{\text{m}}}{2}\!\left(1 + \cos\dfrac{\pi\big(t-(T_\mathrm{acc} + T_\mathrm{plat})\big)}{T_\mathrm{dec}}\right), & T_\mathrm{acc} + T_\mathrm{plat} \le t \le T_{\text{s}},
\end{cases}
\label{eq:vel_program}
\end{equation}
with $U_{\text{m}}\text{=}1$, $T_\mathrm{s}\text{=}T_\mathrm{acc}+T_\mathrm{plat}+T_\mathrm{dec}$, and $T_\mathrm{acc}\text{=}1.0$, $T_\mathrm{plat}\text{=}0.1$, $T_\mathrm{dec}\text{=}2.4$.

\begin{figure}
\centering
\includegraphics[width=\linewidth]{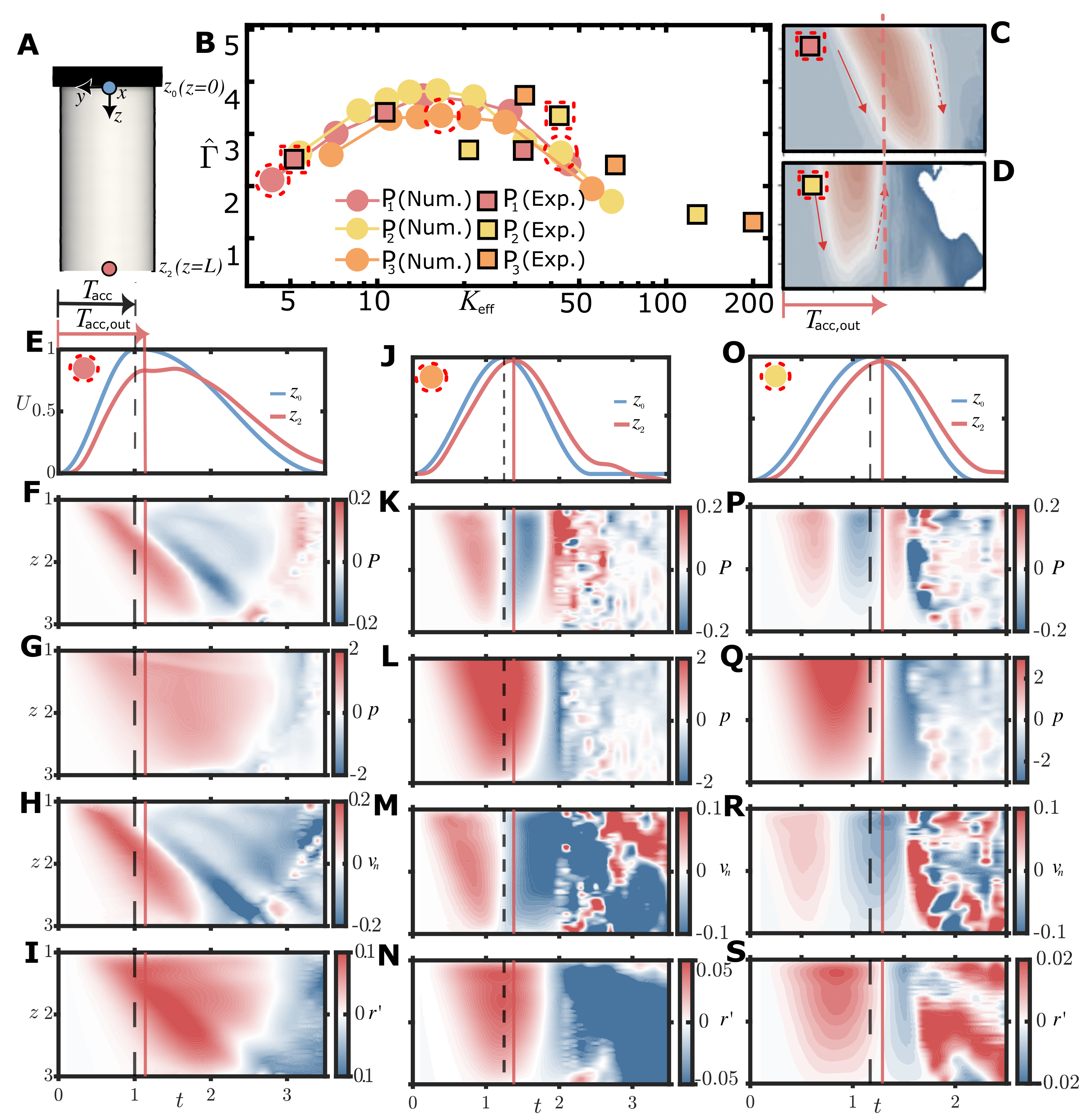}
\caption{\footnotesize \textbf{(A)} Nozzle schematic with two probes on its axis, one at the inlet ($z\text{=}0$) and one at the outlet ($z\text{=}L$).
\textbf{(B)} Circulation ($\hat{\Gamma}$) versus the effective stiffness ($K_{\mathrm{eff}}$) for the three inflow velocity time histories tested (P1--P3, present simulations) together with the experimental data of \cite{choi2024mechanism} (black squares).
Symbols with dashed red marker denote the cases detailed in the following panels.
\textbf{(C-D)} Radial nozzle displacement from~\cite{choi2024mechanism} (their Fig.5) for \textbf{(C)} $K_{\mathrm{eff}}\text{=}10.6$, $\hat{c}\text{=}1.09$, $\tau\text{=}0.48$, and 
\textbf{(D)} $K_{\mathrm{eff}}\text{=}42.65$, $\hat{c}\text{=}3.72$, $\tau\text{=}0.57$.
\textbf{(E--I)} Inflow P1: \textbf{(E)} imposed inflow velocity at the inlet ($z\text{=}0$) and velocity measured at the outlet ($z\text{=}L$) and, along a generatrix of the nozzle in the $(z,t)$ plane: \textbf{(F)} local power \textbf{(G)} pressure on the nozzle, \textbf{(H)} structural-normal velocity and \textbf{(I)} radial displacement.
\textbf{(J--N)} and \textbf{(O--S)} as in \textbf{(E--I)}, for the second (P2) and third (P3) inflow conditions, respectively.}
\label{fig:S1}
\end{figure}

We also tested two further inflow velocity (labeled P2 and P3 in \textcolor{black}{Fig.\ref{fig:S1}\textbf{B}}), obtained by changing the piston acceleration and deceleration phases, $T_\text{acc}$ and $T_\text{dec}$, \textcolor{black}{in Eq.~\ref{eq:vel_program} (P2: $T_\text{acc}\text{=}1.17$, $T_\text{dec}\text{=}1.00$; P3: $T_\text{acc}\text{=}1.24$, $T_\text{dec}\text{=}1.18$)}, so as to reproduce the effective acceleration times $\tau\text{=}0.57$ and $0.64$ of the experiments of \cite{choi2024mechanism}. 

\textcolor{black}{Figure \ref{fig:S1}} shows that the present simulations reproduce the behavior observed experimentally \cite{choi2024mechanism}.
Indeed, at low $K_{\text{eff}}$ both predict a traveling-wave response of the nozzle.
The radial displacement measured by~\cite{choi2024mechanism} (Fig.\ref{fig:S1}\textbf{C} --see their Fig.5 for the color scale) and the present field illustrated in Fig.\ref{fig:S1}\textbf{I} (and obtained with Eq.\ref{eq:vel_program} --denoted as P1 in the legend of Fig.\ref{fig:S1}\textbf{B}) show the expansion--contraction wave propagating along the nozzle without locking to the imposed inflow.
For this case the local power is illustrated in Fig.\ref{fig:S1}\textbf{F}, together with the local pressure in Fig.\ref{fig:S1}\textbf{G} and the structural velocity along the outward-normal in Fig.\ref{fig:S1}\textbf{H}.
At high $K_{\text{eff}}$, the present radial displacement field (Fig.~\ref{fig:S1}\textbf{S}) \textcolor{black}{agrees with the experimental measurements} of \cite{choi2024mechanism} (Fig.~\ref{fig:S1}\textbf{D}), indeed, both show a standing wave response that completes expansion and recoil within $T_\mathrm{acc}$  (i.e. the acceleration time of the imposed inflow) or $T_\mathrm{acc,out}$ (i.e. the acceleration time defined as in \cite{choi2024mechanism} as the time the centerline jet-exit velocity, in the rigid case, reaches peak velocity).
The corresponding local power and its components are illustrated in Fig.\ref{fig:S1}\textbf{P}, Fig.\ref{fig:S1}\textbf{Q} and Fig.\ref{fig:S1}\textbf{R}.
The optimum discussed in the main text is recovered even when the inflow is changed.
Indeed, here the optimum condition is shown for P3 (Fig.\ref{fig:S1}\textbf{J}), where the expansion and recoil of the nozzle match the acceleration and the deceleration phases of the imposed inflow (see Figs.\ref{fig:S1}\textbf{J},\textbf{N}).
The local power and its components are in Figs.\ref{fig:S1}\textbf{K,L,M}, respectively.

\subsection*{\textcolor{black}{Physics-based model -- straight nozzle}}

This section details the simple model introduced in the manuscript.
The goal of the model is to determine the structural stiffness $K^{\mathrm{opt}}$ at which the jet attains maximum hydrodynamic impulse, circulation and exit velocity gains.
The three-dimensional simulations show that, at this $K^{\mathrm{opt}}$, the structural response takes the form of a standing wave, every point of the nozzle dilating and contracting in phase with the piston pulse.
The derivation that follows reproduces this standing-wave signature analytically and casts $K^{\mathrm{opt}}$ in closed form, with no calibration parameter.

The classical Moens--Korteweg coupling between a thin elastic cylindrical shell and the incompressible fluid column it confines supports the propagation of axial fluid--elastic waves \cite{choi2022flow}.
In dimensionless form, the radial perturbation of the nozzle obeys the wave equation
\begin{equation}
\frac{\partial^2 r'}{\partial t^2} - c^2 \frac{\partial^2 r'}{\partial z^2} = 0,
\label{eq:waveeq_supp}
\end{equation}
with characteristic speed $c^2 \text{=} K$ that, in dimensional form reads $c^*\text{=}(E^*h^*/\rho_f^* \ D^*)^{0.5}$.

Seeking modal solutions $r'(z,t) \text{=} \hat R(z)\,e^{-i\omega t}$ and denoting $k\text{=}\omega/c$ as the wavenumber, Eq.~(\ref{eq:waveeq_supp}) reduces to $\partial^2(\hat R)/{\partial z^2} + k^2 \hat R \text{=} 0$, which admits general solution $\hat R(z) \text{=} A\cos(kz) + B\sin(kz)$.
The physical setup fixes the boundary conditions.
At the inlet, the nozzle is anchored to the rigid feeding tube of the mantle, so the structure is clamped, $r'(0,t)\text{=}0$, which selects $A\text{=}0$.
At the outlet the nozzle is unconstrained, $\partial_zr'(L,t)\text{=}0$, which gives $\cos(kL)\text{=}0$ and therefore the discrete spectrum $k_n L \text{=} (2n-1)\pi/2$, $n \text{=} 1, 2, \dots$, with eigenfunctions ($\hat R_n$) and periods ($T_n$),
\begin{equation}
\hat R_n(z) = \sin \left(\frac{(2n-1)\pi z}{2L}\right),
\qquad
T_n = \frac{2\pi}{\omega_n} = \frac{4L}{(2n-1)\ c}.
\label{eq:Tn-supp}
\end{equation}
The fundamental ($n\text{=}1$) is a quarter-wave standing mode with a node at the clamped inlet, an antinode at the free outlet, and period,
\begin{equation}
T_1 = \frac{4L}{c} = \frac{4L}{\sqrt{K}}.
\end{equation}

The structure is driven by the piston acceleration ($\dot U_p$), that consists of two half-sinusoids (\ref{eq:vel_program}) with angular frequencies $\omega_{\text{a}}\text{=}\pi/T_{\text{acc}}$ and $\omega_{\text{d}}\text{=}\pi/T_{\text{dec}}$, respectively.
The two lobes have unequal durations and unequal peak amplitudes, but equal areas in absolute value ($\int \dot U_p dt= \pm U_m$ over the two lobes), so each lobe delivers the same impulse (in magnitude) to the fluid column and, through it, to the elastic structure.
Therefore, we compute the natural frequency of the forcing as the mean of $\omega_a$ and $\omega_d$, as $\omega_p=(\omega_a+\omega_d)/2=\pi(T_{\text{acc}}+T_{\text{dec}})/(2\,T_{\text{acc}}T_{\text{dec}})$, and the corresponding effective forcing period is
\begin{equation*}
T_p=\frac{2\pi}{\omega_p}=\frac{4 T_{\text{acc}}T_{\text{dec}}}{T_{\text{acc}}+T_{\text{dec}}}.
\end{equation*}
The symmetric limit $T_p=2T_{\text{acc}}=2T_{\text{dec}}$ is recovered for $T_{\text{acc}}=T_{\text{dec}}$, and the effective period tends to $T_p\to 4T_\mathrm{acc}$ as $T_\mathrm{dec}\to\infty$, when the deceleration becomes too slow to influence the structure.

Imposing the frequency-matching condition $T_p=T_1$ yields the closed-form stiffness,
\begin{equation}
K^\mathrm{opt} = 16\left(\frac{L}{T_p}\right)^2,
\label{eq:Kopt}
\end{equation}
which depends only on the nozzle length $L$ and the two forcing durations $T_\mathrm{acc}$ and $T_\mathrm{dec}$, with no free parameters.
From a physical point of view, the condition $T_p=T_1$ indicates that the natural period of the structure's fundamental standing mode coincides with the effective duration of the piston pulse, so the structure executes exactly one charge-and-release cycle per pulse, locked in phase with the imposed inflow.
The structure therefore acts as a tuned spring on the jet: during acceleration ($\dot U_p > 0$) the inertial overpressure drives the structure outward, storing elastic energy in the fundamental standing mode, while during deceleration ($\dot U_p < 0$) the structure recoils and releases that energy back to the fluid, reinforcing the leading vortex ring.

\subsection*{Morphometric data}

\begin{figure}[h]
\centering
\includegraphics[width=0.95\linewidth]{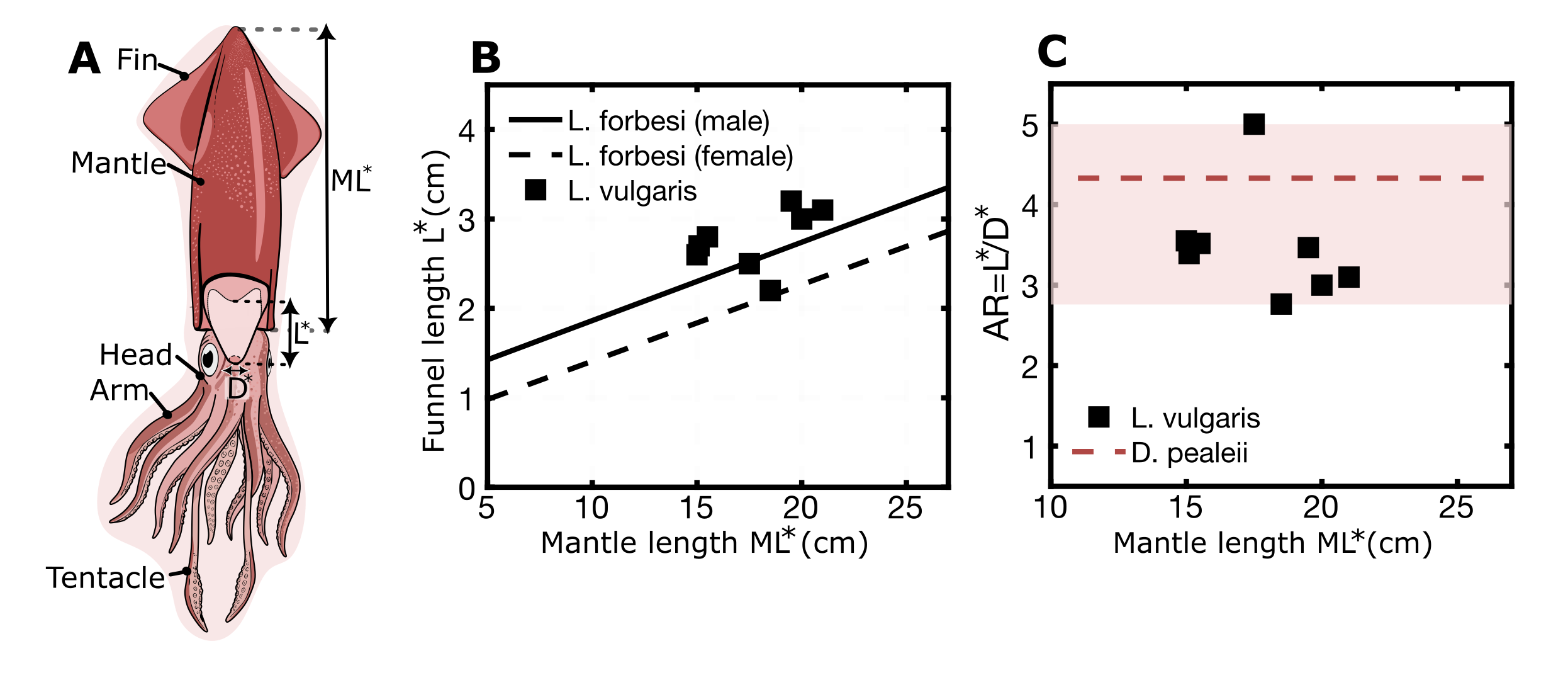}
\caption{\small \textbf{Morphometric data.} (\textbf{A}) Schematic of a squid showing its characteristic features.
(\textbf{B}) Funnel length ($L^*$) versus mantle length ($\text{ML}^*$).
Data for \emph{L.~forbesi} are taken from \cite{emam2014morphometric}.
(\textbf{C}) Ratio of funnel length ($L^*$) to orifice diameter ($D^*$) versus mantle length ($\text{ML}^*$).
Data for \emph{D.~pealeii} are estimated from Fig.1 of \cite{rosenbluth2010muscles} (as individual mantle lengths are not reported, the data point is plotted across the mantle length range given in that study).}
\label{fig:S2}
\end{figure}

Figure~\ref{fig:S2} presents the morphometric data underlying the functional constraints imposed by the curvature discussed in the main text. 
Figure~\ref{fig:S2}\textbf{A} provides a schematic of the squid, highlighting its characteristic lengths.
Figures \ref{fig:S2}\textbf{B} shows funnel length ($L^*$) versus mantle length ($ML^*$) for our own measurements on \emph{L.~vulgaris} and for the dataset of Emam \emph{et al.} \cite{emam2014morphometric} on \emph{L.~forbesi}.
Fig.\ref{fig:S2}\textbf{C} shows the ratio between the funnel length and the funnel diameter ($D^*$) versus mantle length, that remains almost constant across the range of mantle lengths sampled.

\clearpage

\begin{table} 
\centering
\caption{\textbf{Simulation parameters.} Reference scales and dimensionless groups used in the main manuscript. Asterisks denote dimensional quantities, while square brackets indicate the set of values explored.}
\label{tab:S1} 
\begin{tabular}{lll} 
\\
\hline
Category & Variable & Value(s) \\
\hline
Reference & Nozzle diameter, $D^\ast$ & $0.015$\,m \\
		  & Peak inflow velocity, $U^\ast$ & $0.3$\,m\,s$^{-1}$ \\
		  & Fluid density, $\rho_f^\ast$ & $10^3$\,kg\,m$^{-3}$ \\
Flow      & Dynamic viscosity, $\mu^\ast$ & $10^{-3}$\,kg\,m$^{-1}$\,s$^{-1}$ \\
Nozzle    & Nozzle density, $\rho_s^\ast$ & $2\times10^3$\,kg\,m$^{-3}$ \\
		  & Nozzle aspect ratio, $\text{AR}=L^*/D^*$ & $[1.5, 2, 2.5, 3]$ \\
Dimensionless & Reynolds number, $Re$ & $4500$ \\
		          & Nozzle stiffness, $K$ & $[2-\infty]$ \\
		\hline
	\end{tabular}
\end{table}

\end{document}